\documentclass[aps,prl,reprint,superscriptaddress,amsmath,amssymb,floatfix]{revtex4-2}

\usepackage{graphicx}
\usepackage{bm}
\usepackage{hyperref}

\newcommand{\avg}[1]{\langle #1\rangle}
\newcommand{\Li}{\operatorname{Li}}
\newcommand{\z}{\zeta}
\newcommand{\Tc}{T_c}
\newcommand{\dd}{\mathrm{d}}
\newcommand{\Rk}{\mathcal R}
\newcommand{\Tarr}{\mathcal{T}}        
\newcommand{\DT}{\Delta\Tarr}          
\newcommand{\DTcl}{\Delta\Tarr_{\rm cl}}
\newcommand{\runin}[1]{\smallskip\noindent\textbf{#1}\ }

\begin{document}

\title{A universal time-of-arrival signature of Bose--Einstein condensation}

\author{Mathieu Beau}
\affiliation{Department of Physics, University of Massachusetts, Boston, Massachusetts 02125, USA}

\author{Timothey Szczepanski}
\affiliation{Ecole Polytechnique, Palaiseau, France}

\date{\today}

\begin{abstract}
We show that Bose--Einstein condensation produces a cusp in the time-of-arrival (TOA) statistics of a harmonically trapped gas released into free fall. In the semiclassical long time-of-flight regime, with $\epsilon=\sigma_V/\sqrt{2gH}\ll1$, both the mean and standard deviation of the arrival time distribution, which are governed by the longitudinal velocity variance, remain continuous, but acquire a cusp whose one-sided slope ratio is universal within the ideal-gas far-field limit, $\Rk_\infty=2.5556\ldots$, and equals the trapped-gas specific-heat ratio $C(\Tc^-)/C(\Tc^+)$. Finite atom number rounds the cusp and weak interactions perturb it only weakly, leaving a measurable time-domain signature of condensation.

\end{abstract}

\maketitle

\runin{Introduction.} Phase transitions announce themselves through nonanalyticities in thermodynamic functions; at Bose--Einstein condensation the heat capacity of a harmonically trapped gas has a finite discontinuity at $\Tc$.
Such singularities are normally reconstructed from equilibrium response rather than read from a single dynamical measurement.
Time-of-flight, the diagnostic of ultracold gases~\cite{PethickSmith,Dalfovo1999,Boiron06}, is almost always analyzed in the \emph{space} domain, as a density image at a fixed time, yielding temperature, condensate fraction, and coherence.
Its \emph{time}-domain counterpart, the distribution of detection times at a fixed plane, whose random value we write $\Tarr$ (the temperature is denoted $T$), has been far less exploited as a many-body probe, even though time-resolved atom optics such as temporal slits and pulsed detection have long existed~\cite{Steane1995,Szriftgiser1996,Arndt1996} and single-atom--resolving detectors for metastable species make arrival-time histograms directly accessible~\cite{Robert2001,Vassen2012,Bouton2015}.
We show that this time-domain observable images the condensation specific-heat discontinuity as a universal cusp.

The quantum TOA problem, assigning a distribution to the time at which a particle reaches a point, has a long and unresolved history with no universally accepted operator~\cite{Kijowski1974,Allcock1969,GrotRovelliTate1996,Muga2000}.
Operationally accessible formulations based on the probability current~\cite{Beau2024b,Leavens1998}, on quantum-clock and stroboscopic protocols~\cite{Maccone2020,Lloyd2026,Roncallo2023}, and on projective free-evolution sampling~\cite{Beau2025b} nonetheless converge on the same far-field semiclassical predictions, and the dropped--wave-packet case has been worked out in detail~\cite{Beau2024a,Beau2025c,delCampo2009}.
Free fall is a natural arena: the classical arrival time $t_c=\sqrt{2H/g}$ is fixed by geometry while fluctuations around it encode the initial velocity distribution, and matter-wave free fall is now routinely realized both on Earth and in space~\cite{vanZoest2010,Muntinga2013,Aveline2020,Frye2021,Gaaloul2022}.
Very recently the arrival-time statistics of an atomic condensate have been examined in the experimentally demanding short-distance regime within Gross--Pitaevskii theory, with a measurement proposal based on the position-resolved arrival flux~\cite{Naidon2026,Naidon2026b}.

Here we ask a complementary question: how does Bose statistics near condensation imprint itself on the TOA statistics of a trapped gas released into the \emph{far field}?
We consider an ideal Bose gas in thermal equilibrium in an anisotropic harmonic trap, released from height $H$ above a detector plane, in the long time-of-flight semiclassical regime.
The model is intentionally minimal -- interactions, finite-size effects, and beyond-far-field corrections are excluded -- so that the statistical signature is clean; we then quantify each omission and show the signature survives at experimentally relevant scales.

\runin{TOA from the flux of a falling Bose gas.}
Following the flux description of a ballistically expanding trapped Bose gas~\cite{Boiron06}, the longitudinal flux through a detection plane can be written as
\begin{equation}
\avg{\hat J(z,t)}=
\mathcal N\,
\frac{v_z(t)}{\sqrt{1+\omega_z^2t^2}}\,
\rho_{\rm eq,\perp}(\tilde z),
\label{eq:J_def}
\end{equation}
where $\mathcal N$ is fixed by normalization,
\begin{equation}
\tilde z=\frac{z-z_c(t)}{\sqrt{1+\omega_z^2t^2}},
\qquad
z_c(t)=H-\frac12gt^2,
\label{eq:tilde_z}
\end{equation}
and $\rho_{\rm eq,\perp}$ is the transverse-integrated equilibrium density in the trap.
The TOA distribution is the normalized flux magnitude at the detector plane,
\begin{equation}
\Pi(t)=\mathcal{N}\left|\avg{\hat J(z_{\rm det},t)}\right|,
\label{eq:Pi_def}
\end{equation}
where the normalization factor is defined as $\mathcal{N}^{-1} \equiv \int_0^\infty \dd t\,\Pi(t)=1$, 
and its first two moments define the mean arrival time and the arrival-time standard deviation,
\begin{equation}
\avg{\Tarr}=\int_0^\infty \dd t\,t\,\Pi(t),
\qquad
\DT=\sqrt{\int_0^\infty \dd t\,t^2\,\Pi(t)-\avg{\Tarr}^2}.
\label{eq:moments}
\end{equation}

\runin{Semiclassical reduction.}
In the long time-of-flight regime $\omega_z t_c\gg1$ the longitudinal expansion is ballistic and the arrival time follows from the randomized classical trajectory~\cite{Beau2024a,Beau2025c}, in which the wave-packet spreading $\sigma(t)=\sigma\sqrt{1+t^2/\tau^2}\to\sigma_V\,t$ linearizes into an effective random initial velocity $V$ (Supplemental Material~\cite{SM}), $0=H-\tfrac12g\Tarr^2+V\Tarr$, with $\avg V=0$ and variance $\sigma_V^2=\avg{V^2}$.
Expanding in $\varepsilon=\sigma_V/\sqrt{2gH}\ll1$ gives
\begin{equation}
\avg{\Tarr}\simeq t_c+
\frac{\sigma_V^2}{2g\sqrt{2gH}},
\qquad
\DT\simeq \frac{\sigma_V}{g},
\label{eq:semiclassical_moments}
\end{equation}
the gas analogue of the single dropped--wave-packet result of Refs.~\cite{Beau2024a,Beau2025c}, with the quantum velocity spread $\hbar/2m\sigma$ replaced by the statistical spread $\sigma_V$.
Both moments are controlled by the same $\sigma_V^2$; eliminating it gives
\begin{equation}
\avg{\Tarr}-t_c=\frac{\DT^2}{2t_c},
\label{eq:mean_var_relation}
\end{equation}
so the mean shift and the variance carry identical temperature dependence and therefore an identical nonanalyticity at $\Tc$.

\runin{Above $\Tc$: Bose-statistical renormalization.}
In the high-temperature limit $\sigma_V^2=k_BT/m$, giving the ballistic estimates $\avg{\Tarr}_{\rm cl}-t_c=(k_BT/m)/(2g\sqrt{2gH})$ and $\DTcl=g^{-1}\sqrt{k_BT/m}$.
Close to condensation from above, the equilibrium Bose gas is a sum over permutation cycles~\cite{PethickSmith}: the transverse-integrated longitudinal density is a convex mixture of Gaussians, the cycle-$j$ component having variance $\sigma_{V,j}^2=(k_BT/m)/j$ and \emph{population} weight (the integrated area, including the width factor $\propto j^{-1/2}$)
$p_j=z^{\,j}/(j^{3}\Li_3(z))$, where the fugacity is $\ z=e^{\beta\tilde\mu}$, 
so that $\sum_j p_j=1$ and $N\propto\Li_3(z)$.
Averaging the per-cycle variances with these weights yields
\begin{equation}
\sigma_V^2(T)=\frac{k_BT}{m}\,R(T),
\qquad
R(T)=\frac{\Li_{4}(z)}{\Li_{3}(z)},
\label{eq:sigmaV_polylog}
\end{equation}
consistently with the trapped-gas momentum variance $\avg{p_z^2}=mk_BT\,\Li_4(z)/\Li_3(z)$~\cite{PethickSmith,Dalfovo1999}.
For $T>\Tc$ the fugacity is fixed by $\Li_3(z)=\z(3)(\Tc/T)^3$, so that $\DT(T>\Tc)\simeq g^{-1}\sqrt{k_BT/m}\,\sqrt{\Li_4(z)/\Li_3(z)}$, and at $T=\Tc^+$ ($z\to1^-$),
\begin{equation}
\DT(\Tc^+)\simeq
\DTcl(\Tc)\sqrt{\z(4)/\z(3)}
\simeq0.9489\,\DTcl(\Tc).
\label{eq:universal_factor}
\end{equation}
This is a universal Bose-statistical renormalization of the classical TOA fluctuation; by Eq.~\eqref{eq:mean_var_relation} the mean shift is renormalized by the same factor $\z(4)/\z(3)$.

\begin{figure*}[t]
\centering
\IfFileExists{fig_TOA_fluctuations.pdf}{%
\includegraphics[width=\linewidth]{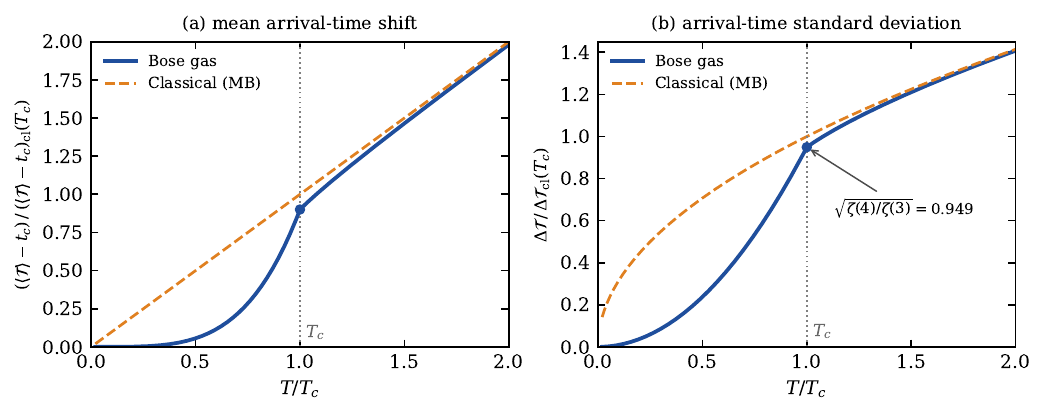}%
}{%
\fbox{\begin{minipage}[c][0.30\linewidth][c]{0.95\linewidth}\centering
Placeholder: mean and std-dev curves vs $T/\Tc$. Filename: \texttt{fig\_TOA\_fluctuations.pdf}
\end{minipage}}%
}
\caption{Time-of-arrival statistics across the Bose--Einstein transition.
(a)~Mean arrival-time shift $\avg{\Tarr}-t_c$ and (b)~standard deviation $\DT$, each normalized to its classical value at $\Tc$, versus $T/\Tc$.
The dashed curves are the Maxwell--Boltzmann predictions, $\avg{\Tarr}_{\rm cl}-t_c\propto T$ and $\DTcl=g^{-1}\sqrt{k_BT/m}$, i.e.\ the values the gas would take in the absence of quantum statistics.
Below $\Tc$ the dashed lines are the (unphysical) continuation of a non-condensing classical gas, shown only as a reference against which the Bose suppression and the cusp are visible.
Both Bose curves are continuous at $\Tc$ but develop a cusp from the onset of macroscopic ground-state occupation; because $\avg{\Tarr}-t_c\propto\sigma_V^2$ while $\DT\propto\sigma_V$, the mean shows a sharper kink than the standard deviation, but the two share the same one-sided slope ratio.
}
\label{fig:TOA_fluctuations}
\end{figure*}

\runin{Below $\Tc$: condensate--thermal mixture.}
Below $\Tc$ the detected signal separates into condensate and thermal contributions 
\begin{equation}\label{Eq:PiDistributionT<Tc}
\Pi(t;T)=w_0(T)\Pi_0(t)+w_T(T)\Pi_T(t)\ ,
\end{equation}
with ideal harmonic-trap weights $w_0(T)=1-(T/\Tc)^3$ and $w_T(T)=(T/\Tc)^3$~\cite{PethickSmith,Dalfovo1999}.
The condensate has velocity variance $\sigma_{V,0}^2=\hbar\omega_z/2m$, while the thermal component has pinned fugacity $z=1$ and $\sigma_{V,T}^2=(k_BT/m)R_1$, $R_1=R(1)=\z(4)/\z(3)$.
To leading semiclassical order the whole-gas moments are
\begin{equation}
\avg{\Tarr}\simeq t_c+
\frac{w_0\sigma_{V,0}^2+w_T\sigma_{V,T}^2}{2g\sqrt{2gH}},
\quad
\DT\simeq
\frac{\sqrt{w_0\sigma_{V,0}^2+w_T\sigma_{V,T}^2}}{g}.
\label{eq:DT_whole}
\end{equation}
Because $w_0$ turns on only for $T<\Tc$, both $\avg{\Tarr}(T)$ and $\DT(T)$ stay continuous while their temperature derivatives jump at the transition, see Fig.~\ref{fig:TOA_fluctuations}.

\runin{Universal cusp.}
The cusp is quantified by the ratio of one-sided slopes, which by Eq.~\eqref{eq:mean_var_relation} is identical for the mean shift and the standard deviation,
\begin{equation}
\Rk_\infty\equiv
\frac{\left.\partial_T\DT\right|_{\Tc^-}}
{\left.\partial_T\DT\right|_{\Tc^+}}
=\frac{\left.\partial_T(\avg{\Tarr}-t_c)\right|_{\Tc^-}}
{\left.\partial_T(\avg{\Tarr}-t_c)\right|_{\Tc^+}}.
\label{eq:R_def}
\end{equation}
With $\hbar\omega_z\ll k_B\Tc$ the condensate variance is negligible at the transition; the below-threshold slope then follows from Eq.~\eqref{eq:DT_whole} ($\sigma_V^2\propto T^4$) and the above-threshold slope from the fugacity derivative of $R(T)$, giving
\begin{equation}
\Rk_\infty=
\frac{4}{1+\Tc R'(\Tc^+)/R_1}
\simeq2.5556,
\label{eq:universal_slope_ratio}
\end{equation}
with $\Tc R'(\Tc^+)/R_1=3[\z(4)\z(2)-\z(3)^2]/[\z(2)\z(4)]\simeq0.5652$.
The result is exact in the joint limit of an ideal gas, harmonic confinement, far-field detection ($\omega_z t_c\gg1$), and the thermodynamic limit; departures from each are controlled and quantified below.
The value is not accidental: by the virial theorem for a harmonic trap $\avg{p_z^2}=mE/3N$, so $\sigma_V^2(T)\propto E(T)$ and the one-sided slope of $\sigma_V^2$ at $\Tc$ is proportional to the heat capacity $C=\partial_T E$.
The kink ratio therefore coincides with the specific-heat ratio of the ideal harmonically trapped gas,
\begin{equation}
\Rk_\infty=\frac{C(\Tc^-)}{C(\Tc^+)}
=\frac{12\,\z(4)/\z(3)}{12\,\z(4)/\z(3)-9\,\z(3)/\z(2)}
\simeq2.5556,
\label{eq:cusp_is_heat_capacity}
\end{equation}
so the TOA cusp is the time-domain image of the condensation specific-heat discontinuity.
In this far-field semiclassical regime the ratio is independent of atom mass, trap frequencies, atom number, and drop height; these set only the overall scale of the TOA moments, not the universal kink ratio.

\begin{figure*}[t]
\centering
\includegraphics[width=\linewidth]{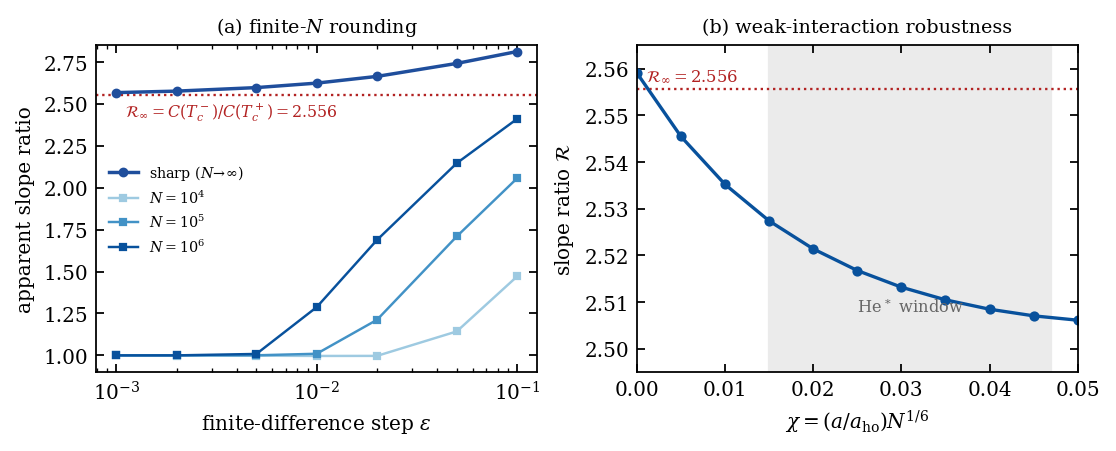}
\caption{Observability and robustness of the TOA cusp.
(a)~Apparent one-sided slope ratio at the transition versus finite-difference temperature step $\varepsilon$.
The sharp thermodynamic-limit curve approaches the universal value $\Rk_\infty\simeq2.5556$ (dotted line), equal to the trapped-gas specific-heat ratio.
The finite-$N$ curves are computed from the exact discrete-level sum for $^4$He ($m=6.64\times10^{-27}$~kg) in a harmonic trap $(\omega_x,\omega_y,\omega_z)=2\pi\times(1000,1000,100)$~Hz, with $N=10^4,10^5,10^6$ (geometric-mean frequency $\bar\omega=2\pi\times464$~Hz, $\Tc=0.45,0.97,2.10\ \mu$K respectively).
At finite $N$, the transition is rounded over a window $\delta T/\Tc\sim N^{-1/3}$; if $\varepsilon$ is pushed below this width, the finite difference samples the smoothed region and the apparent ratio collapses toward unity.
For experimentally realistic atom numbers the measured ratio therefore lies below $\Rk_\infty$ but approaches it from below as $N$ increases.
(b)~Weak-interaction correction to the cusp ratio from the semiclassical Hartree--Fock local-density calculation.
The coupling is $\chi=(a/a_{\rm ho})N^{1/6}$, the dotted line denotes the ideal-gas value $\Rk_\infty$, and the shaded region indicates the dilute metastable-helium window.
Repulsive interactions shift the critical temperature by several percent but change the dimensionless slope ratio only at the percent level, so finite-size rounding is the dominant correction in the far-field regime.
}
\label{fig:observability_robustness}
\end{figure*}

\runin{Experimental scales and limitations.}
The effect is observed through the temperature dependence of the TOA statistics; both the mean arrival time and the width carry the signature, and Eq.~\eqref{eq:mean_var_relation} provides a consistency check between them.
The control parameter is $\varepsilon=\sigma_V/\sqrt{2gH}$, with the universal result holding for $\varepsilon\ll1$ and $\omega_z t_c\gg1$, and the absolute width near $\Tc$ is of order $g^{-1}\sqrt{k_B\Tc/m}$.
Metastable-helium experiments with single-atom--resolving microchannel-plate detection~\cite{Robert2001,Vassen2012,Bouton2015} and long-baseline or microgravity platforms~\cite{vanZoest2010,Muntinga2013,Aveline2020,Frye2021,Gaaloul2022} are natural settings in which $\varepsilon\ll1$ is reached.
Finite atom number rounds the nonanalyticity quantitatively: the transition is smoothed over $\delta T/\Tc\sim N^{-1/3}$, so a measured slope ratio falls below $\Rk_\infty$ and approaches it from below as $N\to\infty$, see Fig.~\ref{fig:observability_robustness}(a); which is a concrete finite-size signature.
The figures use the semiclassical (local-density) form, which becomes exact as $N\to\infty$; the exact discrete-level sum agrees with it to better than $0.1\%$ for $T\ge\Tc$ and merely rounds the cusp below~\cite{SM}.

\runin{Weak-interaction robustness.}
The universal value is an ideal-gas, thermodynamic-limit result, but the cusp is not an artifact of neglecting interactions.
A semiclassical Hartree--Fock calculation (Supplemental Material~\cite{SM}) reproduces the known mean-field downward shift of $\Tc$~\cite{Giorgini1996,Dalfovo1999} and shows that, with coupling $\chi=(a/a_{\rm ho})N^{1/6}$, the below-threshold slope is essentially unchanged while the above-threshold slope rises slightly, so that $\Rk(\chi)\simeq\Rk_\infty(1-0.6\,\chi)$.
Over the range relevant to dilute metastable-helium gases, interactions shift $\Tc$ by several percent but $\Rk$ only at the percent level (about $2\%$ at $\chi=0.05$), see Fig.~\ref{fig:observability_robustness}(b), so finite-$N$ rounding, not weak repulsion, is the dominant correction.
A full interacting treatment including condensate expansion lies beyond the present scope and is complementary to the Gross--Pitaevskii regime of Refs.~\cite{Naidon2026,Naidon2026b}.

\runin{Verification on the metastable-helium platform.}
The signature is within reach of the same metastable-helium setup that underlies the arrival-time proposal of Naidon, Happ, and Boiron~\cite{Naidon2026,Naidon2026b}: a $^4$He$^*$ condensate released onto a single-atom--resolving microchannel-plate detector~\cite{Robert2001,Vassen2012,Bouton2015}.
Whereas that proposal probes the \emph{near}-field, position-resolved arrival flux to discriminate competing TOA operators, the cusp lives in the \emph{far}-field temperature dependence of the same detection record.
For the trap $(\omega_x,\omega_y,\omega_z)=2\pi\times(1000,1000,100)$~Hz with $N=10^6$ atoms ($\Tc=2.1\ \mu$K) and a drop height $H=5$--$50$~cm, the control parameters are $\varepsilon=\sigma_V/\sqrt{2gH}\simeq0.02$--$0.06$ and $\omega_z t_c\simeq60$--$190$, so the far-field semiclassical limit is well satisfied.
The absolute arrival-time width near $\Tc$ is $\DT\simeq\sigma_V/g\simeq6$~ms, comparable to the few-millisecond time-of-flight widths routinely resolved for $^4$He$^*$ condensates~\cite{Robert2001,Bouton2015}, while the mean shift $\avg{\Tarr}-t_c=\DT^2/2t_c\simeq0.1$~ms provides the cross-check of Eq.~\eqref{eq:mean_var_relation}.
The universal slope ratio $\Rk_\infty\simeq2.556$ is rounded over $\delta T/\Tc\sim N^{-1/3}\simeq1\%$ (about $20$~nK at $\Tc=2.1\ \mu$K), within the reach of standard time-of-flight thermometry, and the He$^*$ scattering length $a\simeq7.5$~nm gives $\chi=(a/a_{\rm ho})N^{1/6}\simeq0.03$, hence only a $\sim2\%$ interaction correction [Fig.~\ref{fig:observability_robustness}(b)].
The dominant systematic is therefore finite-$N$ rounding rather than interactions, and the cusp should be resolvable by scanning $T$ across $\Tc$ and extracting the one-sided slopes of $\DT(T)$ on this existing platform.

\runin{Conclusion.}
Bose--Einstein condensation leaves a universal imprint on the time-of-arrival statistics of a freely falling quantum gas.
Above $\Tc$, Bose permutation cycles renormalize the Maxwell--Boltzmann velocity variance by $\Li_4(z)/\Li_3(z)$; below $\Tc$, macroscopic ground-state occupation alters its temperature dependence, so that both the mean arrival time $\avg{\Tarr}(T)$ and the width $\DT(T)$ remain continuous but develop a cusp.
The one-sided slope ratio of that cusp, $\Rk_\infty=C(\Tc^-)/C(\Tc^+)\simeq2.5556$, is independent of atom mass, trap geometry, atom number, and drop height, and equals the trapped-gas specific-heat ratio: a thermodynamic discontinuity read directly from a single time-domain dynamical observable.
The prediction is sharp and falsifiable, finite atom number reduces the measured ratio below $\Rk_\infty$ and weak repulsion shifts it only at the percent level, both in controlled ways, and it is accessible on existing single-atom--resolved metastable-helium and microgravity platforms.
It is complementary to the near-field, interacting-condensate regime and the position-resolved flux proposal of Refs.~\cite{Naidon2026,Naidon2026b}: the same detection record that resolves the spatial arrival flux there encodes, in its far-field temperature dependence, the condensation cusp here.
More broadly, recasting an equilibrium nonanalyticity as a kink in an arrival-time distribution suggests time-domain dynamical probes as a general route to thermodynamic singularities in quantum gases.

\begin{acknowledgments}
\end{acknowledgments}

\bibliographystyle{apsrev4-2}
\bibliography{Ref}

\newpage
\onecolumngrid

\vspace{5mm} 

\begin{center}
\textbf{\large APPENDIX}
\end{center}

This Supplemental Material derives the results quoted in the main text, in the following order: the far-field semiclassical moments of the time of arrival (TOA) from the stochastic representation, by direct analogy with the single-particle results of Refs.~\cite{Beau2024a,Beau2025c}; the longitudinal velocity variance of the trapped ideal Bose gas, with the distinction between the population and density-amplitude weightings; the one-sided slope ratio at the transition; its identification with the specific-heat ratio; the equality of the mean and variance cusps; the finite-$N$ rounding; and the weak-interaction robustness of the cusp. Throughout, $T$ is the temperature, $\Tarr$ the time of arrival, $t_c=\sqrt{2H/g}$ the classical arrival time, and $z=e^{\beta\tilde\mu}$ the fugacity measured from the zero-point energy.

\section{Far-field semiclassical reduction}
\label{sec:reduction}

For a Gaussian system the TOA distribution at a fixed detection point follows from the stochastic representation of Refs.~\cite{Beau2024a,Beau2024b}: writing the measured position at time $t$ as $X_t=x_c(t)+\xi\,\sigma(t)$ with $\xi\sim\mathcal N(0,1)$, the arrival time $\Tarr$ at a fixed plane is the solution of $x=x_c(\Tarr)+\xi\,\sigma(\Tarr)$, and its density equals the normalized probability current there. For a particle dropped from height $H$ with random initial velocity $v_0$, the classical centroid is $x_c(t)=v_0 t+\tfrac12 g t^2$ and the width is $\sigma(t)=\sigma\sqrt{1+t^2/\tau^2}$ with $\tau=2m\sigma^2/\hbar$, so the defining equation reads
\begin{equation}
x=v_0\Tarr+\tfrac12 g\Tarr^2+\xi\,\sigma\sqrt{1+\Tarr^2/\tau^2}.
\label{eq:sm_full}
\end{equation}
In the long time-of-flight (ballistic) regime $\Tarr\gg\tau$ the spreading factor linearizes,
\begin{equation}
\sigma(\Tarr)=\sigma\sqrt{1+\Tarr^2/\tau^2}\;\xrightarrow{\;\Tarr\gg\tau\;}\;\frac{\sigma}{\tau}\,\Tarr=\sigma_v\,\Tarr,
\qquad \sigma_v\equiv\frac{\sigma}{\tau}=\frac{\hbar}{2m\sigma},
\label{eq:sm_linearize}
\end{equation}
so the noise term $\xi\,\sigma(\Tarr)$ becomes \emph{linear} in $\Tarr$ and merges with the drift into an effective random initial velocity $V=v_0+\xi\sigma_v$. Equation~\eqref{eq:sm_full} then collapses to the randomized classical trajectory
\begin{equation}
x=V\Tarr+\tfrac12 g\Tarr^2 ,
\label{eq:sm_quadratic}
\end{equation}
[Ref.~\cite{Beau2024a}, Eq.~(13)]. Writing it for the drop geometry (detector at $x=H$ below the release point, $\avg{v_0}=0$) gives the form used in the main text, $0=H-\tfrac12 g\Tarr^2+V\Tarr$. For the trapped gas, $V$ is simply the initial vertical velocity of an atom drawn from the equilibrium ensemble, with $\avg V=0$ and variance $\sigma_V^2$ (derived in the next section); the single-particle wave-packet spread $\sigma_v$ is thereby replaced by the statistical spread $\sigma_V$. Solving the quadratic and keeping the physical root,
\begin{equation}
\Tarr(V)=\frac{V+\sqrt{V^2+2gH}}{g}
=\frac{V}{g}+t_c\sqrt{1+\frac{V^2}{2gH}}.
\label{eq:sm_TofV}
\end{equation}
For a distribution with $\avg V=0$ and variance $\sigma_V^2=\avg{V^2}$, introduce the small parameter $\varepsilon=\sigma_V/\sqrt{2gH}\ll1$ and expand Eq.~\eqref{eq:sm_TofV} to second order,
\begin{equation}
\Tarr(V)\simeq t_c+\frac{V}{g}+\frac{t_c}{4gH}\,V^2+O(V^3).
\label{eq:sm_expand}
\end{equation}
Taking expectations, and using $t_c/(4gH)=1/(2g\sqrt{2gH})$,
\begin{equation}
\avg{\Tarr}\simeq t_c+\frac{\sigma_V^2}{2g\sqrt{2gH}},
\qquad
\DT^2=\avg{\Tarr^2}-\avg{\Tarr}^2\simeq\frac{\sigma_V^2}{g^2}.
\label{eq:sm_moments}
\end{equation}
Equation~\eqref{eq:sm_moments} is the gas analogue of the single dropped-wave-packet result of Refs.~\cite{Beau2024a,Beau2025c}: there the velocity spread is the quantum value $\sigma_v=\hbar/2m\sigma$ set by the initial width $\sigma$, giving $\DT=t_c\,\sigma_v/\sqrt{2gH}=\sigma_v/g$ and $\avg{\Tarr}-t_c=H\sigma_v^2/(2gH)^{3/2}$ [Ref.~\cite{Beau2024a}, Eqs.~(17)--(18)]; the trapped-gas result follows by the substitution $\sigma_v^2\to\sigma_V^2$, with $\sigma_V^2$ the statistical longitudinal velocity variance derived in the next section. Eliminating $\sigma_V^2$ between the two moments in Eq.~\eqref{eq:sm_moments} gives the model-independent relation
\begin{equation}
\avg{\Tarr}-t_c=\frac{\DT^2}{2t_c},
\label{eq:sm_meanvar}
\end{equation}
used repeatedly below.

\section{Longitudinal velocity variance from the TOA flux}
\label{sec:variance}

We derive the longitudinal velocity variance directly from the plane-integrated
flux formula. This form is useful because the transverse-integrated density
contains cycle amplitudes proportional to $z^{\,j}/j^{5/2}$, whereas the
normalized time-of-arrival distribution is weighted by the time area of each
component. This distinction is the origin of the factor
$\operatorname{Li}_4(z)/\operatorname{Li}_3(z)$.

The plane-integrated current through the detector plane is
\begin{equation}
\langle \hat J(z_{\rm det},t)\rangle
=
\mathcal N
\frac{v_z(t)}{b(t)}
\rho_{\rm eq,\perp}(\tilde z(t)),
\qquad
b(t)=\sqrt{1+\omega_z^2t^2},
\label{eq:sm_flux_start}
\end{equation}
where
\begin{equation}
\tilde z(t)
=
\frac{z_{\rm det}-z_c(t)}{b(t)},
\qquad
z_c(t)=H-\frac12gt^2 .
\label{eq:sm_ztilde}
\end{equation}
The TOA distribution is the normalized flux magnitude,
\begin{equation}
\Pi(t)
=
\left|\langle \hat J(z_{\rm det},t)\rangle\right|,
\qquad
\int_0^\infty \mathrm{d}t\,\Pi(t)=1 .
\label{eq:sm_Pi_def}
\end{equation}

In the semiclassical trap limit $\beta\hbar\omega_\alpha\ll1$, the
transverse-integrated equilibrium density near the transition is
\begin{equation}
\rho_{\rm eq,\perp}(\tilde z)
\simeq
A(T)
\sum_{j=1}^\infty
\frac{z^{\,j}}{j^{5/2}}
\exp\!\left[
-\frac{j m\beta\omega_z^2}{2}\tilde z^2
\right],
\label{eq:sm_rho_perp_sc}
\end{equation}
with
\begin{equation}
z=e^{\beta\tilde\mu},
\qquad
A(T)=
\frac{1}{(\beta\hbar\omega_\perp)^2}
\frac{1}{\sqrt{2\pi\hbar^2\beta/m}} .
\label{eq:sm_AT}
\end{equation}
The factor $z^{\,j}/j^{5/2}$ is the amplitude of the $j$-th longitudinal
Gaussian. It is not the probability weight of that component in the normalized
TOA distribution.

The unnormalized contribution of the $j$-th term to the TOA distribution is
\begin{equation}
\Pi_j^{\rm un}(t)
\propto
\frac{z^{\,j}}{j^{5/2}}
\frac{|v_z(t)|}{b(t)}
\exp\!\left[
-\frac{j m\beta\omega_z^2}{2}\tilde z(t)^2
\right].
\label{eq:sm_Pi_j_un}
\end{equation}
Its weight in the normalized TOA distribution is its time area,
\begin{equation}
I_j
=
\int_0^\infty \mathrm{d}t\,\Pi_j^{\rm un}(t).
\label{eq:sm_Ij_def}
\end{equation}
The flux prefactor is the Jacobian between time and the scaled longitudinal
coordinate:
\begin{equation}
\left|
\frac{\mathrm{d}\tilde z}{\mathrm{d}t}
\right|
=
\frac{|v_z(t)|}{b(t)},
\qquad
\frac{|v_z(t)|}{b(t)}\,\mathrm{d}t
=
|\mathrm{d}\tilde z|.
\label{eq:sm_jacobian}
\end{equation}
Therefore
\begin{align}
I_j
&\propto
\frac{z^{\,j}}{j^{5/2}}
\int \mathrm{d}\tilde z\,
\exp\!\left[
-\frac{j m\beta\omega_z^2}{2}\tilde z^2
\right]
\nonumber\\
&\propto
\frac{z^{\,j}}{j^{5/2}}
\frac{1}{\sqrt j}
=
\frac{z^{\,j}}{j^3}.
\label{eq:sm_Ij}
\end{align}
Thus the normalized weight of the $j$-th component is
\begin{equation}
p_j
=
\frac{I_j}{\sum_{n\ge1} I_n}
=
\frac{z^{\,j}/j^3}{\operatorname{Li}_3(z)}.
\label{eq:sm_pj_correct}
\end{equation}

This is the key point. If one incorrectly normalized the amplitudes
$z^{\,j}/j^{5/2}$ directly, one would use
\begin{equation}
p_j^{\rm amp}
=
\frac{z^{\,j}/j^{5/2}}{\operatorname{Li}_{5/2}(z)},
\label{eq:sm_pj_wrong}
\end{equation}
which weights the height of each Gaussian instead of its time area.

We now identify the velocity variance associated with each component. In the
long time-of-flight regime,
\begin{equation}
b(t)\simeq \omega_z t,
\label{eq:sm_ballistic_b}
\end{equation}
and the arrival condition may be written as the randomized classical equation
\begin{equation}
0
=
H-\frac12g\mathcal T^2+V\mathcal T .
\label{eq:sm_random_classical}
\end{equation}
Equivalently, in the far field, the scaled coordinate is proportional to the
initial vertical velocity,
\begin{equation}
\tilde z
\simeq
\frac{V}{\omega_z},
\label{eq:sm_ztilde_velocity}
\end{equation}
up to an irrelevant sign convention. Substituting this into the Gaussian factor
of Eq.~\eqref{eq:sm_Pi_j_un} gives
\begin{equation}
\exp\!\left[
-\frac{j m\beta\omega_z^2}{2}\tilde z^2
\right]
=
\exp\!\left[
-\frac{j m\beta}{2}V^2
\right].
\label{eq:sm_velocity_gaussian}
\end{equation}
Hence the $j$-th component has longitudinal velocity variance
\begin{equation}
\sigma_{V,j}^2
=
\frac{1}{jm\beta}
=
\frac{k_BT}{jm}.
\label{eq:sm_sigmaV_j}
\end{equation}

The whole-cloud velocity variance is the area-weighted average over the
normalized TOA components:
\begin{align}
\sigma_V^2
&=
\sum_{j\ge1}
p_j\,\sigma_{V,j}^2
\nonumber\\
&=
\sum_{j\ge1}
\frac{z^{\,j}/j^3}{\operatorname{Li}_3(z)}
\frac{k_BT}{jm}
\nonumber\\
&=
\frac{k_BT}{m}
\frac{\sum_{j\ge1} z^{\,j}/j^4}
{\sum_{j\ge1} z^{\,j}/j^3}.
\label{eq:sm_sigmaV_sum}
\end{align}
Therefore
\begin{equation}
\boxed{
\sigma_V^2(T>T_c)
=
\frac{k_BT}{m}
\frac{\operatorname{Li}_4(z)}{\operatorname{Li}_3(z)}
}.
\label{eq:sm_sigmaV_above}
\end{equation}
The fugacity above the transition is fixed by the harmonic-trap number
constraint
\begin{equation}
\operatorname{Li}_3(z)
=
\zeta(3)
\left(
\frac{T_c}{T}
\right)^3 .
\label{eq:sm_fugacity_constraint}
\end{equation}

Below the transition, the fugacity is pinned at $z=1$. The thermal component
has
\begin{equation}
\sigma_{V,T}^2
=
\frac{k_BT}{m}
\frac{\zeta(4)}{\zeta(3)}
\equiv
\frac{k_BT}{m}R_1,
\qquad
R_1=\frac{\zeta(4)}{\zeta(3)}.
\label{eq:sm_sigmaVT_below}
\end{equation}
Only the excited fraction carries this thermal variance:
\begin{equation}
w_T(T)
=
\frac{N_{\rm exc}}{N}
=
\left(
\frac{T}{T_c}
\right)^3,
\qquad
w_0(T)
=
1-
\left(
\frac{T}{T_c}
\right)^3 .
\label{eq:sm_weights_below}
\end{equation}
The condensate contributes the harmonic-oscillator zero-point velocity
variance
\begin{equation}
\sigma_{V,0}^2
=
\frac{\hbar\omega_z}{2m}.
\label{eq:sm_sigmaV0}
\end{equation}
Thus the whole-cloud variance below the transition is
\begin{equation}
\sigma_V^2(T<T_c)
=
w_0(T)\frac{\hbar\omega_z}{2m}
+
w_T(T)\frac{k_BT}{m}R_1 .
\label{eq:sm_sigmaV_below_full}
\end{equation}
In the semiclassical thermodynamic regime $\hbar\omega_z\ll k_BT_c$, the
condensate zero-point contribution is negligible at the transition and does not
affect the leading one-sided slope. Hence, near $T_c$,
\begin{equation}
\boxed{
\sigma_V^2(T<T_c)
\simeq
\frac{k_BT}{m}R_1
\left(
\frac{T}{T_c}
\right)^3
=
\frac{k_B R_1}{mT_c^3}T^4
}.
\label{eq:sm_sigmaV_below}
\end{equation}

Finally, the far-field randomized classical equation gives
\begin{equation}
\langle \mathcal T\rangle
\simeq
t_c+
\frac{\sigma_V^2}{2g\sqrt{2gH}},
\qquad
\Delta \mathcal T
\simeq
\frac{\sigma_V}{g}.
\label{eq:sm_TOA_moments_from_sigmaV}
\end{equation}
Thus the Bose-statistical factor
$\operatorname{Li}_4(z)/\operatorname{Li}_3(z)$ controls both the mean TOA
shift and the TOA width.

\section{One-sided slope ratio at the transition}
\label{sec:cusp}

Because $\DT=\sqrt{\sigma_V^2}/g$ is continuous at $\Tc$, the ratio of one-sided derivatives of $\DT$ equals the ratio of one-sided derivatives of $\sigma_V^2$,
\begin{equation}
\Rk_\infty
=\frac{\partial_T\DT|_{\Tc^-}}{\partial_T\DT|_{\Tc^+}}
=\frac{\partial_T\sigma_V^2|_{\Tc^-}}{\partial_T\sigma_V^2|_{\Tc^+}} .
\label{eq:sm_ratio_def}
\end{equation}
From Eq.~\eqref{eq:sm_sigmaV_below}, $\sigma_V^2\propto T^4$ below the transition, so (in units $k_B=m=\Tc=1$)
\begin{equation}
\partial_T\sigma_V^2\big|_{\Tc^-}=4R_1=\frac{4\,\z(4)}{\z(3)} .
\label{eq:sm_slope_below}
\end{equation}
Above the transition $\sigma_V^2=T\,R(T)$ with $R=\Li_4(z)/\Li_3(z)$ and $z(T)$ fixed by $\Li_3(z)=\z(3)\,T^{-3}$. Differentiating the constraint and using $\Li_s'(z)=\Li_{s-1}(z)/z$,
\begin{equation}
\frac{\Li_2(z)}{z}\frac{\dd z}{\dd T}=-3\z(3)\,T^{-4}
\;\Longrightarrow\;
\frac{\dd z}{\dd T}\Big|_{\Tc^+}=-\frac{3\z(3)}{\z(2)} ,
\label{eq:sm_dzdT}
\end{equation}
while
\begin{equation}
\frac{\dd R}{\dd z}\Big|_{z=1}
=\frac{1}{z}\frac{\Li_3^2-\Li_4\Li_2}{\Li_3^2}\Big|_{z=1}
=\frac{\z(3)^2-\z(4)\z(2)}{\z(3)^2} .
\label{eq:sm_dRdz}
\end{equation}
Hence
\begin{equation}
\Tc R'(\Tc^+)=\frac{\dd R}{\dd z}\Big|_{1}\frac{\dd z}{\dd T}\Big|_{\Tc^+}
=\frac{3\big[\z(4)\z(2)-\z(3)^2\big]}{\z(3)\z(2)},
\qquad
\frac{\Tc R'(\Tc^+)}{R_1}
=\frac{3\big[\z(4)\z(2)-\z(3)^2\big]}{\z(2)\z(4)}\simeq0.5652 ,
\label{eq:sm_TcRp}
\end{equation}
and $\partial_T\sigma_V^2|_{\Tc^+}=R_1+\Tc R'(\Tc^+)$. Combining with Eq.~\eqref{eq:sm_slope_below},
\begin{equation}
\boxed{\;
\Rk_\infty=\frac{4R_1}{R_1+\Tc R'(\Tc^+)}
=\frac{4}{1+\Tc R'(\Tc^+)/R_1}
=\frac{4\,\z(4)}{4\,\z(4)-3\,\z(3)^2/\z(2)}
\simeq2.5556 \; }
\label{eq:sm_Rinf}
\end{equation}
The result is independent of $m$, $\bar\omega$, $H$ and $N$, since these enter only the overall scale of $\sigma_V^2$ and cancel in the ratio.

\section{Identification with the specific-heat ratio}
\label{sec:heat}

The trapped ideal gas obeys the virial theorem $\avg{\rm KE}=\avg{V_{\rm tr}}=E/2$, so the total kinetic energy is $E/2$ and, by isotropy of the momentum distribution, $\avg{p_z^2}=\tfrac13\avg{p^2}=mE/3N$. Therefore
\begin{equation}
\sigma_V^2=\frac{\avg{p_z^2}}{m^2}=\frac{E}{3Nm},
\qquad
\partial_T\sigma_V^2=\frac{C}{3Nm},\quad C=\partial_T E .
\label{eq:sm_virial}
\end{equation}
The one-sided slopes of $\sigma_V^2$ are thus proportional to the one-sided heat capacities, and Eq.~\eqref{eq:sm_ratio_def} gives $\Rk_\infty=C(\Tc^-)/C(\Tc^+)$. Using $E=3Nk_BT\,\Li_4(z)/\Li_3(z)$ above and $E=3Nk_B[\z(4)/\z(3)]\,T^4/\Tc^3$ below~\cite{PethickSmith,Dalfovo1999},
\begin{equation}
\frac{C(\Tc^-)}{Nk_B}=12\frac{\z(4)}{\z(3)}\simeq10.80,
\qquad
\frac{C(\Tc^+)}{Nk_B}=12\frac{\z(4)}{\z(3)}-9\frac{\z(3)}{\z(2)}\simeq4.23,
\label{eq:sm_C}
\end{equation}
so that
\begin{equation}
\Rk_\infty=\frac{C(\Tc^-)}{C(\Tc^+)}
=\frac{12\,\z(4)/\z(3)}{12\,\z(4)/\z(3)-9\,\z(3)/\z(2)}\simeq2.5556 ,
\label{eq:sm_Cratio}
\end{equation}
in exact agreement with Eq.~\eqref{eq:sm_Rinf}. The TOA cusp is therefore the time-domain image of the well-known specific-heat discontinuity of the trapped Bose gas.

\section{Equality of the mean and variance cusps}
\label{sec:meanvar}

By Eq.~\eqref{eq:sm_meanvar}, $\avg{\Tarr}-t_c=\DT^2/(2t_c)$ with $t_c$ a temperature-independent constant; both the mean shift and the variance are proportional to $\sigma_V^2(T)$. Their one-sided derivatives at $\Tc$ therefore share the same ratio,
\begin{equation}
\frac{\partial_T(\avg{\Tarr}-t_c)|_{\Tc^-}}{\partial_T(\avg{\Tarr}-t_c)|_{\Tc^+}}
=\frac{\partial_T\sigma_V^2|_{\Tc^-}}{\partial_T\sigma_V^2|_{\Tc^+}}
=\Rk_\infty .
\label{eq:sm_mean_ratio}
\end{equation}
The standard deviation $\DT=\sqrt{\sigma_V^2}/g$ has the same ratio because $\sqrt{\sigma_V^2}$ is continuous at $\Tc$. The mean shift, being linear in $\sigma_V^2$, exhibits a visually sharper kink than $\DT$, which is linear in $\sigma_V$; the slope \emph{ratio} is identical.

\begin{figure}[h!t]
\centering
{\includegraphics[width=0.6\linewidth]{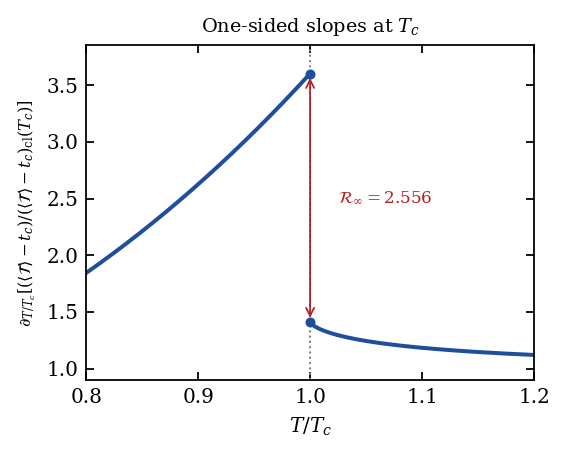}}{}
\caption{One-sided slopes of the normalized mean arrival-time shift,
$\partial_{T/\Tc}\big[(\avg{\Tarr}-t_c)/(\avg{\Tarr}-t_c)_{\rm cl}(\Tc)\big]$, across the transition.
The slope drops from $4\,\z(4)/\z(3)\simeq3.60$ at $\Tc^-$ to $R_1+\Tc R'(\Tc^+)\simeq1.41$ at $\Tc^+$, a one-sided ratio $\Rk_\infty\simeq2.556$ equal to the trapped-gas specific-heat ratio.
The standard deviation $\DT$ shares the same ratio but a visually milder kink, being linear in $\sigma_V$ rather than $\sigma_V^2$.}
\label{fig:cusp}
\end{figure}

Figure~\ref{fig:cusp} shows the one-sided slopes of the normalized mean arrival-time shift at $\Tc$: the below-threshold slope $4\,\z(4)/\z(3)\simeq3.60$ exceeds the above-threshold slope $R_1+\Tc R'(\Tc^+)\simeq1.41$ by exactly the universal factor $\Rk_\infty\simeq2.556$.

\section{Finite-$N$ rounding}
\label{sec:finiteN}

At finite $N$ the condensate fraction turns on smoothly rather than as a nonanalytic kink. The transition is rounded over a temperature window $\delta T/\Tc\sim N^{-1/3}$ set by the spacing between the ground and first excited trap levels relative to $k_B\Tc$. Consequently a numerically or experimentally estimated one-sided slope ratio depends on both $N$ and the temperature step $\varepsilon$ used to evaluate it. If $\varepsilon$ is taken smaller than the rounding width $N^{-1/3}$, the one-sided differences both sample the smoothed region and the apparent ratio drops toward unity; if $\varepsilon$ straddles the rounding width, the sharp-limit slopes are recovered. The universal value $\Rk_\infty$ is reached only in the joint limit $N\to\infty$ followed by $\varepsilon\to0$, i.e.\ by extrapolating $N\to\infty$ at fixed $\varepsilon$ that exceeds the rounding width [Fig.~\ref{fig:observability_robustness}(a) of the main text]. This is a genuine finite-size effect rather than a numerical artifact: a real experiment at finite $N$ measures a slope ratio below $\Rk_\infty$, approaching it as $N$ increases.

\begin{figure}[t]
\centering
{\includegraphics[width=0.7\linewidth]{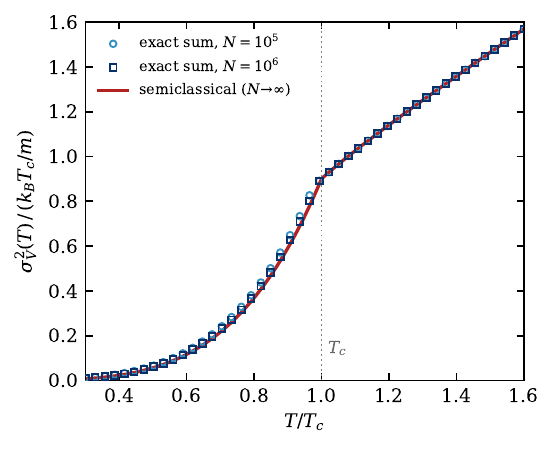}}{}
\caption{Longitudinal velocity variance $\sigma_V^2(T)$ from the exact discrete-level sum [Eq.~\eqref{eq:sm_exact}, open symbols] versus the semiclassical thermodynamic-limit form (solid), for $^4$He ($m=6.64\times10^{-27}$~kg) in the harmonic trap $(\omega_x,\omega_y,\omega_z)=2\pi\times(1000,1000,100)$~Hz ($\bar\omega=2\pi\times464$~Hz), at $N=10^5$ ($\Tc=0.97\ \mu$K) and $N=10^6$ ($\Tc=2.10\ \mu$K). The two agree to $<0.1\%$ for $T\ge\Tc$; the exact sum rounds the cusp over $\delta T/\Tc\sim N^{-1/3}$ and the below-$\Tc$ excess decreases with $N$.}
\label{fig:exact_vs_sc}
\end{figure}

The semiclassical (local-density) expressions used in Eqs.~\eqref{eq:sm_sigmaV_above}--\eqref{eq:sm_sigmaV_below} and in the main-text figures are not an additional approximation beyond the thermodynamic limit: they become exact as $N\to\infty$, which is the same limit in which the cusp sharpens into a true nonanalyticity. To verify this, we evaluated $\sigma_V^2(T)$ from the \emph{exact} discrete-level sum for the anisotropic trap $(\omega_x,\omega_y,\omega_z)=2\pi(1000,1000,100)$~Hz, resummed as the cycle (cluster) series
\begin{equation}
N=\sum_{k\ge1}\frac{z^{\,k}}{\prod_i(1-e^{-k\beta\hbar\omega_i})},
\quad
\avg{p_z^2}=\frac{\hbar m\omega_z}{N}\sum_{k\ge1}\frac{z^{\,k}\coth(k\beta\hbar\omega_z/2)}{2\,\prod_i(1-e^{-k\beta\hbar\omega_i})},
\label{eq:sm_exact}
\end{equation}
which includes the ground state, zero-point motion (the condensate variance $\hbar\omega_z/2m$ arises from the $k\to\infty$ tail), anisotropy, and finite $N$ with no continuum approximation. As shown in Fig.~\ref{fig:exact_vs_sc}, the exact sum coincides with the semiclassical curve to better than $0.1\%$ for $T\ge\Tc$ even at $N=10^5$; below $\Tc$ it lies a few percent above the sharp curve (the residual condensate and finite-size contributions), the excess shrinking as $N$ grows. The finite-$N$ curves of Fig.~\ref{fig:observability_robustness}(a) of the main text are computed from the same exact sum, Eq.~\eqref{eq:sm_exact}, with each one-sided ratio evaluated about the finite-$N$ transition temperature (located as the inflection of the condensate fraction). Since the cusp ratio is built from the one-sided slopes at $\Tc$ and the above-threshold branch is essentially exact, the semiclassical treatment captures $\Rk_\infty$ faithfully. Finally, finite size shifts the transition itself; for the present trap the leading shift is~\cite{Dalfovo1999} $\delta\Tc/\Tc\simeq-\tfrac{\z(2)}{2\z(3)^{2/3}}(\bar\omega_a/\bar\omega)N^{-1/3}\simeq-1.10\,N^{-1/3}$, i.e.\ $-2.4\%$ at $N=10^5$ and $-1.1\%$ at $N=10^6$, with $\bar\omega_a$ the arithmetic and $\bar\omega$ the geometric mean trap frequency.

\section{Weak-interaction robustness of the cusp}
\label{sec:interactions}
 
The universal ratio $\Rk_\infty=C(\Tc^-)/C(\Tc^+)=2.5556\ldots$ of the main text is an ideal-gas result. Here we estimate the leading correction from weak repulsive interactions within a semiclassical Hartree--Fock local-density approximation (HF--LDA)~\cite{Dalfovo1999}. The aim is not to solve the interacting TOA problem but to test whether the dimensionless cusp survives the mean field near the transition. Because the longitudinal momentum distribution stays isotropic in the local-density approximation, only the geometric-mean frequency $\bar\omega$ enters the dimensionless ratio, and an isotropic trap may be used.
 
\paragraph{Model.}
The thermal cloud is described by a local Bose distribution with fugacity $z(\xi)=e^{-\alpha(\xi)}$, where $\xi$ is the dimensionless radius defined by $\beta V_{\rm ho}=\xi^2$. The thermal atoms move in the Hartree--Fock mean field $2g\,n_T$ (direct plus exchange), so the local fugacity obeys the self-consistent equation
\begin{equation}
\alpha(\xi)=\xi^2-\bar\mu+\kappa(\theta)\,\Li_{3/2}\!\big(e^{-\alpha(\xi)}\big),
\qquad \theta=\frac{T}{\Tc^{0}},
\label{eq:sm_hf}
\end{equation}
with $\Tc^{0}$ the ideal trapped-gas critical temperature, $\bar\mu=\mu/k_BT$, and
\begin{equation}
\kappa(\theta)=\kappa_c\sqrt{\theta},
\qquad
\kappa_c=\frac{4\chi}{\sqrt{2\pi}\,\z(3)^{1/6}},
\qquad
\chi=\frac{a}{a_{\rm ho}}N^{1/6}.
\label{eq:sm_kappa}
\end{equation}
Above $\Tc$ the chemical potential is fixed by the atom number,
\begin{equation}
\theta^{3}\!\int_0^\infty\!\xi^2\,\Li_{3/2}\!\big(e^{-\alpha(\xi)}\big)\,\dd\xi=\frac{\sqrt\pi}{4}\,\z(3).
\label{eq:sm_number}
\end{equation}
At and just below $\Tc$ the thermal cloud is saturated at the trap center, $z(0)=1$ ($\alpha(0)=0$), which by Eq.~\eqref{eq:sm_hf} fixes
\begin{equation}
\bar\mu=\kappa(\theta)\,\z(3/2).
\label{eq:sm_pin}
\end{equation}
The condensate absorbs the remaining atoms. We retain only the thermal-cloud contribution to the velocity variance: the direct condensate contribution turns on as $(\Tc-T)^{7/5}$ and is longitudinally suppressed by the trap anisotropy, so it does not generate the leading one-sided slope at $\Tc$. The normalized variance is
\begin{equation}
\frac{\sigma_V^2}{k_B\Tc^{0}/m}=\frac{4\,\theta^4}{\sqrt\pi\,\z(3)}\int_0^\infty\!\xi^2\,\Li_{5/2}\!\big(e^{-\alpha(\xi)}\big)\,\dd\xi,
\label{eq:sm_sv2}
\end{equation}
and, since $\avg{\Tarr}-t_c\propto\sigma_V^2$ and $\DT\propto\sigma_V$, the one-sided slope ratio $\Rk(\chi)=s_-/s_+$ with $s_\pm=\partial_\theta(\sigma_V^2/[k_B\Tc^0/m])|_{\theta_c^\pm}$ controls both observables.
 
\paragraph{Numerics.}
Because $\Li_{3/2}$ has a square-root branch point at $z=1$, where the cusp lives, we evaluate the Bose functions in $\alpha=-\ln z$ using the Robinson expansion $\Li_s(e^{-\alpha})=\Gamma(1-s)\,\alpha^{\,s-1}+\sum_{k\ge0}\z(s-k)(-\alpha)^k/k!$ near $\alpha=0$ and the convergent series elsewhere, and solve Eq.~\eqref{eq:sm_hf} by monotone bisection rather than fixed-point iteration. In the ideal limit the scheme returns the analytic benchmarks $\Rk_\infty=2.5556$, $\z(4)/\z(3)=0.9004$, $s_-(0)=4\z(4)/\z(3)=3.6016$, and $s_+(0)=1.4093$ to better than $0.1\%$, and it reproduces the mean-field critical-temperature shift $\delta\Tc/\Tc^{0}\simeq-1.25\,\chi$, close to the Hartree--Fock value $-1.33\,\chi$ of Ref.~\cite{Giorgini1996}.
 
\begin{figure}[t]
\centering
{\includegraphics[width=0.6\linewidth]{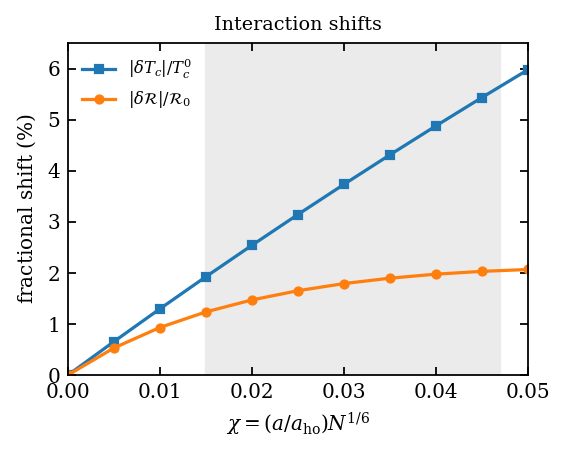}}{}
\caption{Weak-interaction robustness from the HF--LDA calculation. The fractional shift of $\Rk$ stays at the percent level while $\Tc$ shifts several times more strongly.}
\label{fig:interactions}
\end{figure}

\paragraph{Result.}
Repulsive interactions lower $\Tc$ by several percent over the weak-coupling interval, whereas the dimensionless slope ratio changes only at the percent level (Fig.~\ref{fig:interactions}). The below-threshold slope is essentially interaction independent and the above-threshold slope rises slightly, giving
\begin{equation}
\Rk(\chi)\simeq\Rk_\infty\,(1-0.6\,\chi),
\qquad
\Rk(0.02)\simeq2.52,\quad \Rk(0.05)\simeq2.51,
\end{equation}
while $\Tc/\Tc^{0}$ falls by about $2.5\%$ at $\chi=0.02$ and $6\%$ at $\chi=0.05$. Thus weak mean-field repulsion shifts the absolute transition temperature far more than the normalized TOA kink ratio, supporting the interpretation of $\Rk_\infty$ as a robust dynamical signature, with finite-$N$ rounding the dominant correction in the far-field regime. This estimate is a thermodynamic-limit, mean-field result that keeps the thermal--thermal interaction; the condensate back-action on the thermal cloud is narrow near $\Tc$ and $1/N$-suppressed there, and the beyond-mean-field critical window is negligibly narrow in a trap. A complete account including condensate hydrodynamics is complementary to the Gross--Pitaevskii regime of Refs.~\cite{Naidon2026,Naidon2026b}.
 
\end{document}